\documentclass[review]{elsarticle}

\usepackage{hyperref}
\journal{Engineering Applications of Artificial Intelligence}

\usepackage{amsmath}
\usepackage{multirow}
\usepackage{enumitem}  
\usepackage{amsmath}
\usepackage{amssymb}
\usepackage{subcaption}

\usepackage{booktabs}

\newcommand{\proofofref}{}
\newproof{zproofof}{Proof of \proofofref}

\begin{document}

\begin{frontmatter}

% \title{Elsevier \LaTeX\ template\tnoteref{mytitlenote}}
\title{Active Foundational Models for Fault Diagnosis of Electrical Motors}
% \tnotetext[mytitlenote]{Fully documented templates are available in the elsarticle package on \href{http://www.ctan.org/tex-archive/macros/latex/contrib/elsarticle}{CTAN}.}

% \author{Authors}
% \address{Radarweg 29, Amsterdam}
% \fntext[myfootnote]{Since 1880.}

% \author[mymainaddress,mysecondaryaddress]{Elsevier Inc}

% \author[mysecondaryaddress]{Global Customer Service\corref{mycorrespondingauthor}}
% \cortext[mycorrespondingauthor]{Corresponding author}

% \address[mymainaddress]{1600 John F Kennedy Boulevard, Philadelphia}
% \address[mysecondaryaddress]{360 Park Avenue South, New York}

\author[IITM-AM]{Sriram Anbalagan}
\author[IITM-EE]{Sai Shashank GP}
\author[KSU]{Deepesh Agarwal}
\author[KSU]{Balasubramaniam Natarajan}
\author[IITM-AM]{Babji Srinivasan\corref{CorrAuth}}

\address[IITM-AM]{Department of Applied Mechanics and Biomedical Engineering, IIT Madras, Chennai, Tamil Nadu 600036, India}
\address[IITM-EE]{Department of Electrical Engineering, IIT Madras, Chennai, Tamil Nadu 600036, India}
\address[KSU]{Department of Electrical and Computer Engineering, Kansas State University, Manhattan, Kansas 66506, USA}

\cortext[CorrAuth]{Corresponding author. Email:babji.srinivasan@iitm.ac.in}

\begin{abstract}
% This template helps you to create a properly formatted \LaTeX\ manuscript.
Fault detection and diagnosis of electrical motors are of utmost importance in ensuring the safe and reliable operation of several industrial systems. Detection and diagnosis of faults at the incipient stage allows corrective actions to be taken in order to reduce the severity of faults. The existing data-driven deep learning approaches for machine fault diagnosis rely extensively on huge amounts of labeled samples, where annotations are expensive and time-consuming. However, a major portion of unlabeled condition monitoring data is not exploited in the training process. To overcome this limitation, we propose a foundational model-based Active Learning framework that utilizes less amount of labeled samples, which are most informative and harnesses a large amount of available unlabeled data by effectively combining Active Learning and Contrastive Self-Supervised Learning techniques. It consists of a transformer network-based backbone model trained using an advanced nearest-neighbor contrastive self-supervised learning method. This approach empowers the backbone to learn improved representations of samples derived from raw, unlabeled vibration data. Subsequently, the backbone can undergo fine-tuning to address a range of downstream tasks, both within the same machines and across different machines. The effectiveness of the proposed methodology has been assessed through the fine-tuning of the backbone for multiple target tasks using three distinct machine-bearing fault datasets. The experimental evaluation demonstrates a superior performance as compared to existing state-of-the-art fault diagnosis methods with less amount of labeled data. 
\end{abstract}


\end{frontmatter}

\section{Introduction}

Electrical motors are the backbone of various industries, ensuring reliable and efficient operation across various applications, ranging from commercial and industrial settings to aerospace, computer systems, robotics, and defence. Despite their versatility and reliability, electrical motors are susceptible to various faults that can disrupt performance, jeopardize system safety, and lead to costly downtime. Timely and efficient fault detection in electrical motors is paramount to maintaining optimal functionality, reducing maintenance expenses, and preventing catastrophic failures. 

Contemporary approaches for fault diagnosis of electrical motors encompass analytical models and data-driven methodologies. The analytical and physics-based model relies on signal processing techniques to extract domain-invariant features shared among various faults, thereby developing the motor fault diagnosis model \cite{agarwal2020fault}, \cite{agarwal2020ipms} \cite{shen2021physics} \cite{anbalagan2023physics}. Nonetheless, a limitation of this approach arises from the challenge of identifying universal physical features applicable to all fault types.
In contrast, data-driven models leverage machine learning (ML) and deep learning (DL) techniques, often requiring less expertise than their analytical counterparts. Traditional ML methods, including Artificial Neural Networks (ANN), Support Vector Machines (SVM) \cite{gryllias2012support}, Random Forest (RF) \cite{roy2020autocorrelation}, and k-nearest Neighbor (k-NN) \cite{triguero2019transforming}, are widely used for fault detection. These ML approaches come into play following signal processing to extract features from raw vibration data. Notably, these ML methods are considered shallow networks and may encounter difficulties in effectively extracting fault-related features from raw vibration data in complex environments.

Deep learning constitutes a pivotal branch of machine learning (ML), distinguished by its superior implicit feature extraction compared to shallower machine learning methods. The efficacy of deep learning is particularly notable as the quantity of labeled data scales up. Within the realm of intelligent fault diagnosis, various deep learning methods, such as Convolutional Neural Networks (CNN), Recurrent Neural Networks (RNN), and Long Short-Term Memory (LSTM), have recently emerged as focal points of research \cite{zhao2020intelligent} \cite{liu2018fault} \cite{yang2018rotating}.
One notable development in deep learning is the Transformer model, a technique centring on attention mechanisms (AM) and widely employed in diverse research domains \cite{vaswani2017attention}. The uniqueness of the transformer lies in its capacity to assign greater importance to crucial features while extracting domain-discriminative features from input data, all through an attention mechanism. This innovative model is founded on a novel encoder-decoder architecture, primarily relying on the multi-head self-attention mechanism and feed-forward neural networks. It is designed to tackle sequence-to-sequence tasks with an increased ability to capture global dependencies \cite{du2022fault}. In contrast to alternative networks like convolutional and recurrent networks, Transformer-based models demonstrate superior performance in various computer vision applications \cite{dai2019transformer}. Recent work in \cite{du2022fault} introduced a fault diagnosis approach that leverages a transformer with an optimized stacked autoencoder for analysing rotating machinery.
However, even with these advancements, many deep learning algorithms for motor fault diagnosis often rely heavily on abundant labeled samples and the assumption that training and testing data originate from the same distribution, which is not valid in practical scenarios. Acquiring fault data under real-world conditions can be challenging and may involve extended machine operation under fault conditions, which is often impractical. Consequently, gathering an ample number of labeled samples can be problematic. Deep learning models can struggle to establish effective decision boundaries when labelled samples are scarce, resulting in suboptimal performance.
Despite collecting substantial amounts of motor condition monitoring data, manual annotation remains a costly, error-prone, and labor-intensive process. This frequently results in a significant portion of data remaining underutilized due to either the absence of meaningful labels or insufficient samples for specific fault cases, particularly in scenarios with limited labeled data. In such circumstances, supervised learning approaches often perform inadequately. Existing semi-supervised methods struggle to address the challenge of fault detection with extremely limited labeled samples \cite{rombach2021contrastive} \cite{razavi2018semi}.
Self-supervised learning can effectively overcome this limitation, efficiently extracting meaningful features from raw, unlabeled data, thus eliminating the need for expensive labels \cite{chen2020simple}. Currently, numerous self-supervised learning methods have been developed, primarily in the domain of image analysis. These methods involve enabling the network to learn visual features by constructing pretext tasks, such as predicting image rotation, image coloring, random cropping, and more.
However, mechanical vibration signals differ significantly from image data. They are characterized by non-stationary patterns and strong periodicity, necessitating unique pretext tasks tailored to these distinctive properties. Popular signal transformation techniques are commonly employed to define pretext tasks, including random zeroing, random increase, random decrease, time disorder, and random jittering \cite{wang2022self}. Encoders trained to predict these pretext tasks are expected to learn general features that prove valuable for downstream tasks that typically require expensive annotations.
Another prominent category of self-supervised learning techniques utilizes contrastive losses across various computer vision applications \cite{schroff2015facenet}. These methods aim to create a latent space that brings positive samples closer together (e.g., adjacent segments from the same time series signal) while pushing apart negative samples (e.g., segments from different time series signals). Recently, a variant of contrastive learning known as instance discrimination has exhibited remarkable performance in various downstream tasks \cite{chen2020simple}. In this instance discrimination paradigm, the latest work has introduced the concept of non-trivial positives among augmented samples from the same image and even across different images \cite{dwibedi2021little}. Instead of focusing on clustering or prototype learning, they maintain a support set of image embeddings and employ nearest neighbors from this set to define positive samples.
Taking inspiration from these advancements, we have introduced the instance discriminative technique with the nearest neighbor from the support set as a positive sample to the field of electrical motor fault diagnosis for the first time. This pioneering approach has demonstrated a significant improvement over existing self-supervised techniques, marking a promising advancement in this domain.

In all the tasks related to fault classification, obtaining labels for data is often laborious and expensive, while a vast pool of unlabeled data remains readily accessible. Furthermore, training datasets frequently contain redundant samples, leading to protracted classifier training without commensurate improvements in classification performance. To address these challenges, the development of Active Learning (AL) techniques has aimed to pinpoint the most valuable samples for manual labeling, ultimately streamlining the classifier training process \cite{yin2019incorporate}.
AL has emerged as a pivotal concept, promising to reduce annotation costs. It typically encompasses iterative procedures in which the learning algorithm gains autonomy in selecting additional training data. One of the pupular query stratgies in AL is uncertainty sampling, which is designed to solicit labels for samples characterized by high uncertainty. This uncertainty can be quantified through posterior probabilities and marginal entropy, as various strategies for identifying uncertain samples are explored in the literature \cite{chen2015bearing}. Subsequently, these selected informative samples are manually labeled and seamlessly integrated into the training process, enhancing the classifier's performance.
While active learning (AL) is recognized as a supervised approach incurring higher annotation costs than self-supervised learning (SSL), this study critically evaluates the advantages and limitations of both AL and SSL methodologies. As a result, this paper introduces an innovative fault diagnosis approach for industrial equipment that harmoniously integrates SSL with AL. 

The key contributions of this work are as follows:
\begin{enumerate}
    \item 
    In this study, we have introduced a novel foundational model-based AL framework for electrical motor fault diagnosis. This model consists of two key components: the construction of the backbone model, followed by the fine-tuning procedure for various target tasks using less amount of labeled samples.

    \item Instead of solely relying on data augmentation techniques, we have employed an advanced method called instance discrimination-based contrastive Self-Supervised Learning (SSL) to train the transformer-based backbone model. In contrastive self-supervised learning, we have harnessed the benefits of using the nearest neighbor as a positive sample. This approach allows the transformer encoder to learn a more robust representation of the vibration signal, resulting in a well-generalized backbone model.

    \item We have implemented AL strategy based on entropy and Kullback-Leibler (KL) divergence as combined heuristics for strategic sample selection. This helps in identification of most informative samples for annotation within the iterative training loop. The experimental results demonstrate that the proposed model can learn effective decision boundaries even with a limited number of labeled samples. This achievement overcomes the laborious and costly task of manual labeling.

    \item Our research amalgamates the strengths of Active Learning (AL) and the nearest neighbor contrastive SSL method. We have proposed a framework for constructing the backbone model, enabling it to learn more refined representations of the data with minimum labeled samples. Consequently, the proposed model is highly effective in various downstream tasks within the same machine as well as across different machines. The proposed approach has been thoroughly evaluated using three publicly available bearing fault datasets. 

    The remainder of this paper is organized as follows. Section \ref{sec:prelim} briefly provides background on Fault Diagnosis of Electrical Motors, Transformers, Contrastive Learning, Active Learning, and Foundational Models. The components of the proposed active foundational model for fault diagnosis of electrical motors are elaborated in Section \ref{sec:methodology}. Section \ref{sec:results} evaluates the performance of the proposed model over three experimental datasets, and concluding remarks are presented in Section \ref{sec:conclusion}.

\end{enumerate}

\section{Preliminaries} \label{sec:prelim}

\subsection{Fault Diagnosis of Electrical Motors}

An electrical motor comprisess of mechanical elements such as the stator, rotor, bearings and electrical components like windings and end rings. They are designed to operate across various industrial and environmental conditions, subjecting them to diverse forms of stress. These stress factors lead to different types of faults within the electrical motor, broadly categorized as mechanical or electrical faults. Among the faults that afflict electrical motors, bearing faults, shaft misalignment, rotor imbalance, and inter-turn short circuit faults are most prominent. Existing literature indicates that most electrical motor breakdowns can be attributed to mechanical faults \cite{choudhary2019condition}. In this paper, we delve into addressing the primary mechanical faults commonly observed in electrical motors. This includes bearing faults of various sizes occurring at different locations (such as inner-race, ball, and outer-race) and shaft misalignment faults exhibiting varying degrees of severity. 

\subsection{Transformers}
The multi-head self-attention mechanism of the transformer network enables it to capture better the complex dynamics of input data. The key components of transformer architecture are described as follows:
\subsubsection{Mulihead Self-Attention}
The Multi-head attention gives the transformer greater power to encode multiple relationships and nuances for each input sample \cite {guo2022attention}. A mapping from a query ($Q$) and a collection of key-value pairs ($K$) to an output ($V$), where $Q, K$, and $V$ are all vectors, can be thought of as an attention mechanism. The weight matrix (Attention Distribution, AD) is derived from $Q$ and $K$. The result is a weighted sum over $V$. Characterizing sequence similarity is the essence of attention distribution. The attention relationship is mainly calculated by the scaled dot-product attention method. Given a set of input matrices, $ Q \in \mathbb{R}^{m \times n}, K \in \mathbb{R}^{m \times n}$, and $V \in \mathbb{R}^{m \times n}$, the mathematical expression of self-attention calculation is given as follows \cite{wen2022transformers}:

\begin{equation}
\text{Attention}(Q,K,V) = \text{softmax} \left( \frac{QK^{T}}{\sqrt{n}} \right)V 
\end{equation}

\noindent where $\sqrt{n}$ is a scale factor.

The multi-head self-attention enables parallel computation of self-attention by getting multiple sets of $Q, K$, and $V$ matrices in different initialization forms. The concurrent computation of several sets of attention relations is then translated into a set of outputs using a transformation matrix $W^{0}$. It can be expressed as \cite{wen2022transformers}:

\begin{equation}
\begin{aligned}
    \text{MultiHead}(Q,K,V) = Concat(head_{1},..., head_{i})W^{o}\\ 
     head_{i} = \text{Attention} (Q\mathrm{W}_{i}^{Q}, K\mathrm{W}_{i}^{K}, V\mathrm{W}_{i}^{V} )
\end{aligned}
\end{equation}

\subsubsection{Transformer block}:
The transformer block includes layer normalization (LN), feedforward neural network (FNN), and residual connections (RC) along with the above-discussed MSA layer. A transformer block is the composition of transformer encoders and transformer decoders. Given an input of $ \text{x} \in \mathbb{R}^{m \times n} $, the mathematical expression of LN is given as follows:

\begin{equation}
    \text{LN}(x) = \frac{x-\mu}{\sqrt{\sigma^{2}+\varepsilon}} \cdot g+b
\end{equation}

where $\mu$ and $\sigma^{2}$ represent the mean and variance of x. $g$ and $b$ are scaling parameters and translation parameters, respectively; and $\varepsilon$ generally takes the value of 1$e^{-5}$.\\

The attention relations calculated parallel in the feed-forward network can be mathematically expressed as follows: 
\begin{equation}
    \text{FNN}(x)= \text{(GeLU}(x)\;W_{1}+b_{1})\;W_{2}+b_{2}
\end{equation}
where GeLU (Gaussian error Linear Unit) is a nonlinear activation function, $W_{1}$ and $W_{2}$ indicate weights, and $b_{1}$ and $b_{2}$ indicate biases. Considering $x$ as input, $f(x)$ is the output of the weight layer. The calculation of residual connections is expressed as follows:

\begin{equation}
    \text{RC}(x) = f(x)+x.
\end{equation}

Hence, for a transformer block, the output $x_{o}$ of its complete forward propagation process can be represented as \cite{cen2023mask}:
\begin{equation}
\begin{aligned}
    \hat{x} = \text{LN}(\text{MSA}(x_{i}))+ x_{i}\\
     x_{o}= \text{LN}(\text{FNN}(\hat{x}))+\hat{x}
\end{aligned}
\end{equation}

\subsection{Contrastive Learning}
\label{sec:prelim_cl}
Contrastive Learning is a paradigm of SSL based on the idea that samples of the same class are closer to each other in latent representation space and samples of different classes are farther. Contrastive Learning has found many applications in Computer Vision and Pattern Recognition. Learning in the Contrastive Learning framework is achieved by pushing ``similar" looking samples towards each other and ``dissimilar" looking samples farther from each other. \cite{chen2020simple} provides a loss function called \textit{NT-Xent}, which achieves the above-mentioned Learning task via a Contrastive Learning framework. The loss function is given as, 
\begin{equation*} \label{eqn:CLoss}
    l(i, j) = -\log{{\exp\left({\text{sim}\left({z_i}, {z_j}\right)/{\tau}}\right)}\over{\sum_{k=1}^{2N} {\mathbb{I}_{[k \neq i]}}\exp({\text{sim}\left({z_i}, {z_k}\right)/{\tau})}}}
\end{equation*}
Where $\text{sim(u, v)}$ is defined as, 
\begin{equation*}
    \text{sim}(u, v) = {{\langle u, v \rangle}\over{{\lVert u \rVert}{\lVert v \rVert}}}
\end{equation*}
The loss functions employ the projections of the samples $i, j$ onto the latent space, which are ${z_i}, {z_j}$ and the similarity between the two projections are represented using $\text{sim}(u, v)$ and ${\tau}$ is the softmax temperature. This similarity metric is used to determine how ``similar" $i, j$ are with respect to the other samples in the batch of size $2N$.  The loss is high if $i, j$ are more ``similar" compared to other samples in the batch and vice versa. The framework aims to learn the representation space where ``similar" samples are closer. 

\subsection{Active Learning}
AL comprises of machine learning methods that aim to enhance the performance of the decision models by iteratively requesting labels for strategically selected unlabeled samples in the dataset from an external source of labeling, such as domain experts (i.e., Oracle) \cite{agarwal2021addressing}. The key idea revolves around enabling the active learner to engage and interact with the Oracle in order to acquire labels for the most informative examples. AL heuristics like uncertainty, entropy, KL divergence and classification margin are employed to choose the most informative samples from the unlabeled pool of data. This approach leads to improved performance using much fewer labeled training instances compared to traditional supervised machine learning techniques \cite{agarwal2021addressing2}. The utilization of such semi-supervised learning approach proves to be highly advantageous in various applications that make use of modern machine learning pipelines. This is particularly relevant in cases where the process of annotating data might be costly, challenging, or time-intensive.

\subsection{Foundational Models}

Transfer learning and scale provide the technical background for foundational models \cite{bommasani2021opportunities}. Transfer learning involves the application of acquired knowledge from one task to another. Pretraining is the predominant approach to implementing transfer learning in the field of deep learning. The model undergoes training on a surrogate task and is subsequently tuned to align with the downstream task of interest. The scalable attribute enhances the potential of the foundational model. In recent times, there has been a notable adoption of foundational models in the domains of natural language processing (NLP) and computer vision. An example of a task in Bidirectional Encoder Representations from Transformers (BERT) \cite{devlin2018bert} is the masked language modelling challenge, which entails predicting a word that is absent from a phrase by considering its surrounding context. Nevertheless, the profound influence of these fundamental models on NLP does not merely stem from their ability to generate text but rather from their exceptional versatility and adaptability. We have implemented the concept of foundational models in our recent work on fault diagnosis of electrical machines \cite{anbalagan2023foundational}. In this paper, we extend the idea of foundational models to incorporate AL and contrastive SSL so that the transformer-based backbone network can be trained using fewer labeled instances as compared to the state-of-the-art. 

\section{Methodology} \label{sec:methodology}
\begin{figure}
    \centering
    \includegraphics[width=\linewidth, height=250pt ]{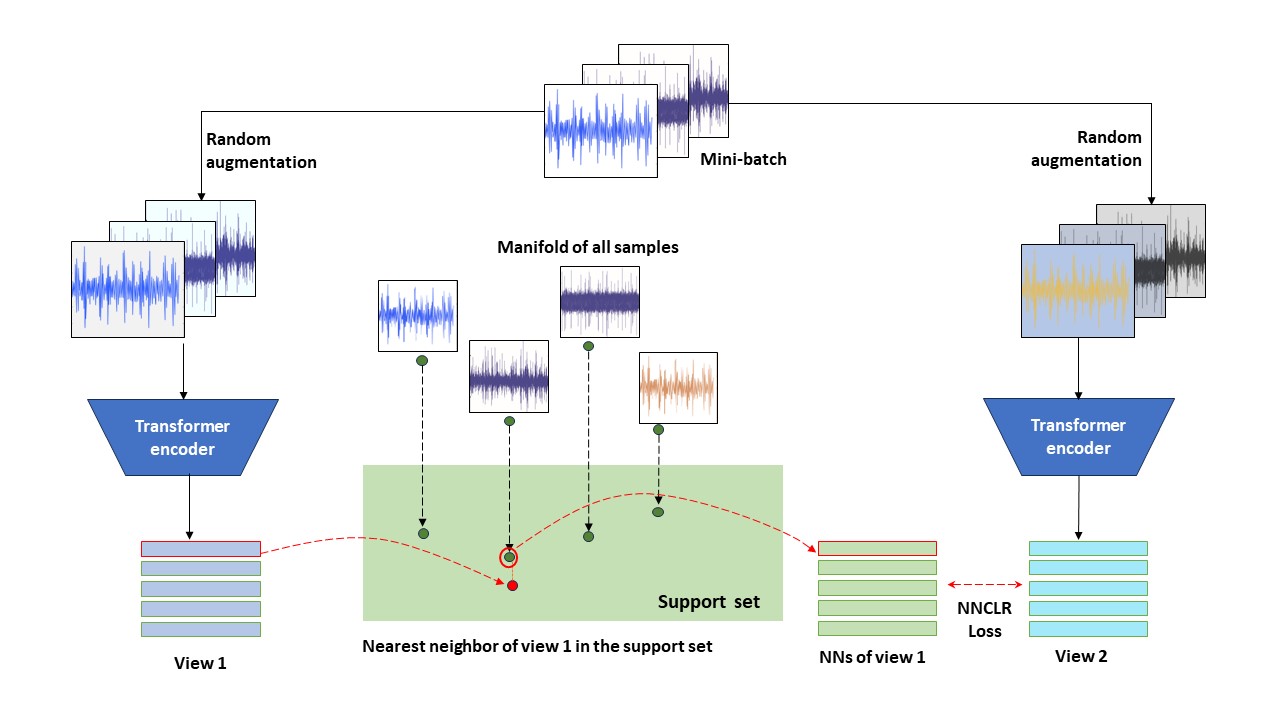}
    \caption{NNCLR training for developing active foundational models for fault diagnosis of electrical motors.}
    \label{fig:nnclrtraining}
\end{figure}

In this paper, we propose an active foundational model for fault detection and diagnosis that can be employed for target tasks across different electrical motors. This is achieved by effectively combining the aspects of AL and contrastive SSL. The proposed model $\mathcal{M}$ consists of two components, a backbone encoder $\mathcal{B}$ and a feed-forward neural network $\mathcal{N}$. The backbone training framework is primarily inspired by Nearest Neighbor - Contrastive Learning Representation framework or NNCLR \cite{dwibedi2021little}, where contrastive Learning framework is utilized, as discussed in Section \ref{sec:prelim_cl}, with slight modifications in terms of robustness to a wide range of semantic augmentations. The implementation of the NNCLR framework in the context of 1-dimensional vibration signal input is schematically illustrated in Figure \ref{fig:nnclrtraining}. This framework replaces one of its augmentations with its nearest neighbor from the dynamic feature space set $\mathcal{S}$ during the forward pass of the training loop. The adoption of nearest neighbors as positive samples in the contrastive loss function defined in Section \ref{eqn:CLoss} results in a modified loss function as shown below. 
\begin{equation*}
    l(i, j) = -\log{{\exp\left({\text{sim}\left(\text{NN}({z_i}, \mathcal{S}), {z_j}\right)/{\tau}}\right)}\over{\sum_{k=1}^{n} {\mathbb{I}_{[k \neq i]}}\exp({\text{sim}\left(\text{NN}({z_i}, \mathcal{S}), {z_k}\right)/{\tau})}}}
\end{equation*}
where,
\begin{equation*}
    \text{NN}(z, \mathcal{S}) = \arg \min_{s \in \mathcal{S}} {\lVert z-s \rVert} 
\end{equation*}

\begin{figure} 
    \centering
    \begin{center}
        \begin{subfigure}{0.5\textwidth}
        \includegraphics[width=1.44\linewidth, height = 240pt]{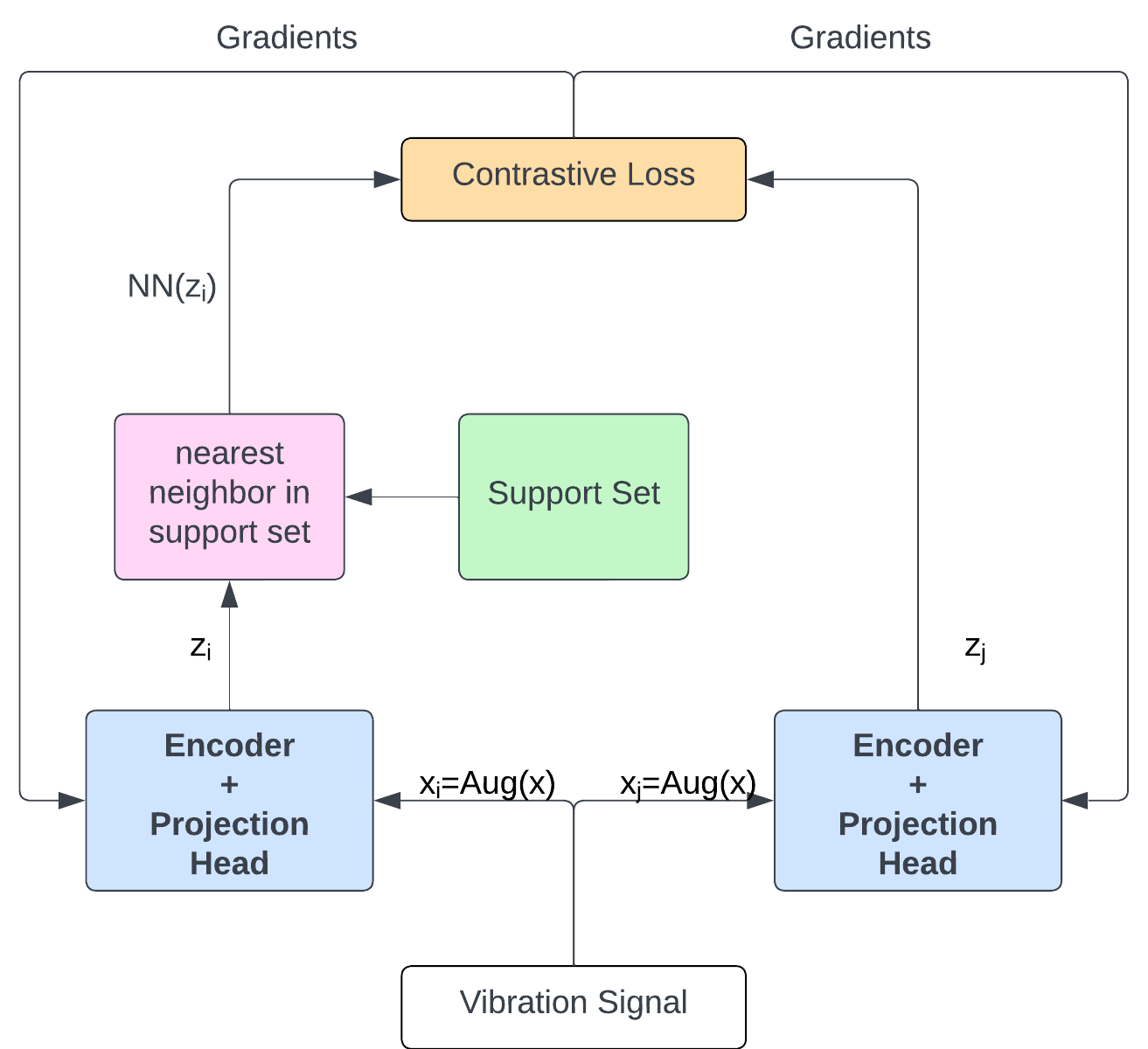}
        \caption{Backbone Training in NNCLR Framework}
        \label{fig:nnclr1}
    \end{subfigure}
    \end{center}
    \par\bigskip
    \begin{center}
        \begin{subfigure}{0.5\textwidth}
        \includegraphics[width=1.44\linewidth, height = 200pt]{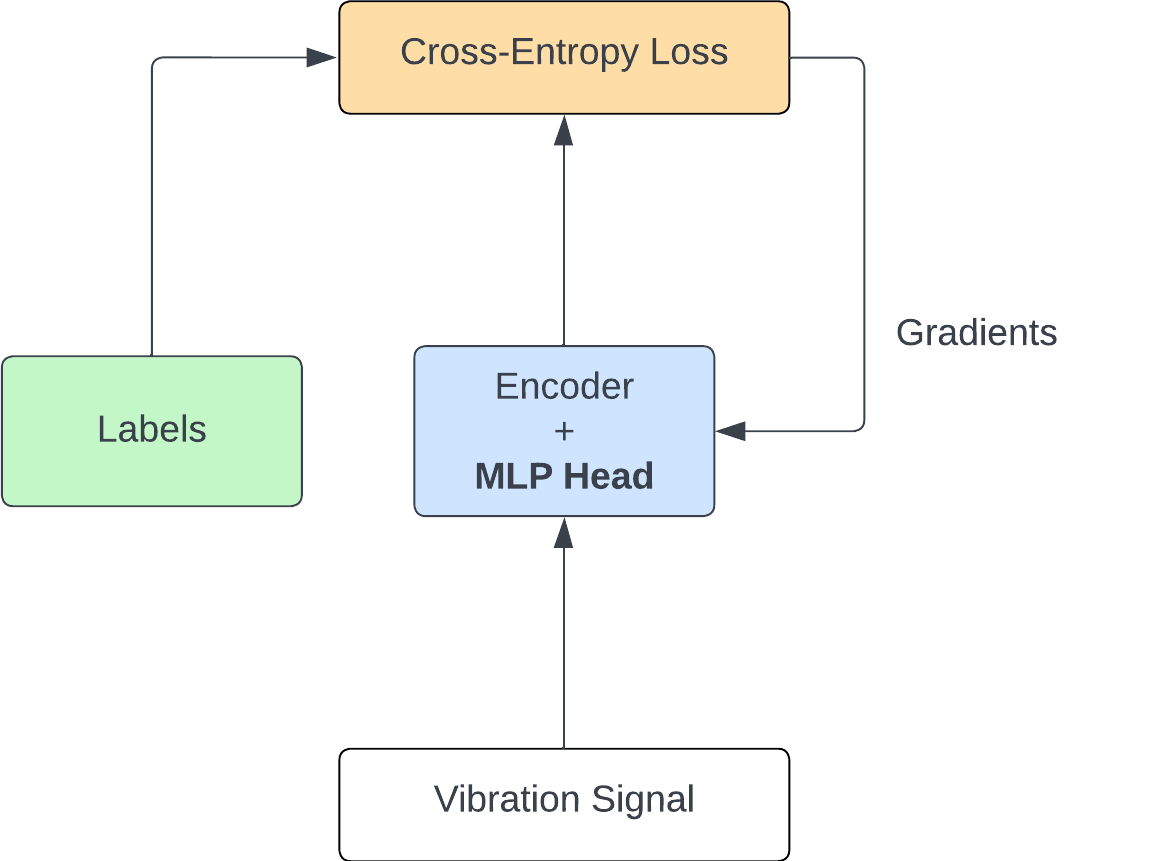}
        \caption{Feed Forward Network Training in NNCLR Framework}
        \label{fig:nnclr2}
    \end{subfigure}
    \end{center}
    \caption{Training Flowcharts of $\mathcal{B}$ and $\mathcal{N}$. Bold indicates that Gradients are not frozen and BackProp is enabled}
    \label{fig:nnclr-flowchart}
\end{figure}

The objective of the Contrastive Learning framework is to learn the representation space where ``similar" samples are closer. The training of Backbone $\mathcal{B}$ and Feed Forward Network $\mathcal{N}$ takes place separately where the $\mathcal{B}$ learns the representation space of similar samples and the $\mathcal{N}$ draws better decision boundaries. The flowchart of the proposed training method is demonstrated in Figure \ref{fig:nnclr-flowchart}. More details on the training setup are provided in Section \ref{subsec: exp and imp}.

Once $\mathcal{B}$ and $\mathcal{N}$ are trained appropriately, their performance is improved by iteratively querying the most informative samples from the unlabeled pool of data by AL within a semi-supervised setting. Entropy of class probabilities and KL-Divergence \cite{10.1214/aoms/1177729694} between actual probability distribution to be predicted and the predicted probability distribution by $\mathcal{M}$ for the particular sample are used as AL heuristics for the semi-supervised training. The training of $\mathcal{B}$ using the mentioned AL heuristics is detailed in Section \ref{subsubsec:bbtrain}.

\subsection{NNCLR framework} \label{subsec:nnclrfw}
Training $\mathcal{B}$ and $\mathcal{N}$ in the NNCLR framework according to Figure \ref{fig:nnclr-flowchart} requires the following terminology to be defined. 
\begin{itemize}
    \item \textit{Contrastive Augmenter:} a signal augmentation module that employs robust data augmentations on the input signals, guided by the hyperparameters specified for contrastive augmentation.
\item \textit{Classification Augmenter:} a signal augmentation module that utilizes mild data augmentations on the input signals, following the hyperparameters designated for classification augmentation.
    \item \textit{Transformer Encoder:} an encoder network based on the Transformer architecture, tasked with extracting feature embeddings of a high-dimensional nature from the augmented signals.
    \item \textit{Projection Head:} a non-linear multi-layer perceptron (MLP) that maps the feature embeddings to a lower dimensional space.
    \item \textit{Linear Probe:} a single dense layer that maps the feature embeddings to the logits for the number of output classes specific to the dataset being trained.
    
\end{itemize}

In addition to the aforementioned SSL training blocks in the architecture, NNCLR \cite{dwibedi2021little} demonstrates the advantage of using nearest neighbors as positive samples in contrastive self-supervised learning. We used this approach as it has been proved to be superior to an augmented view embedding method like SimCLR \cite{chen2020simple} for contrastive learning. Data enrichment is essential for contrastive self-supervised learning systems to operate optimally. However, NNCLR is less reliant on intricate data augmentation methods because the samples from the closest neighbors already have a wide range of variations. Therefore, we employed a few data augmentations along with the nearest neighbors as positive samples. The steps involved in data augmentation and obtaining the nearest neighbors are discussed as follows:     

\begin{enumerate}
    \item \textit{Data Augmentation:}
In this work, we used two techniques for data augmentation: random jittering and random zeros. Random jittering is achieved by introducing random noise to the time series data. Typically, this noise is taken from a random distribution with a known mean and variance. The variety and uncertainty in machine data operating in varying circumstances can be simulated by adding random noise. Random zeros involve randomly adding zeros to time series data, which entails randomly zeroing out parts or segments. This data augmentation method is beneficial for simulating real-world situations where some information may be missing or unreliable by performing data dropout. A probability or percentage that expresses the chance of zeroing out a data point or segment can be used to start the random zero procedure. The frequency of missing data points is controlled by a probability parameter.

\item \textit{Support set:} In order to get the nearest neighbors, NNCLR framework utilizes a support set that stores the embeddings of a subset of the dataset in the memory. The support set is continuously replaced during the training. Increasing the size of the support set helps in improving overall performance. The likelihood of approaching the nearest neighbor from the entire dataset improves when an extensive support set is used. This variability in the positives, not addressed by pre-defined data augmentation strategies, is likely the cause of the performance gain. 
During training, the support set is implemented as a MoCo-like queue \cite{he2020momentum}. In contrastive losses, the nearest neighbors in the queue are employed to locate positives. 
\end{enumerate}

Random augmentations, like random increases, decreases, or zeros, cannot provide positive pairs for various views, deformations of the same signal, or even other identical instances within a semantic class. The data augmentation pipeline is heavily responsible for generalization but cannot account for all variations within a particular class. About the instance discrimination task, NNCLR proposes to move beyond single instance positives. Doing this allows us to develop features more resistant to various views, deformations, and variations within a single class.

\subsection{Implementation Details} \label{subsec: exp and imp}

A time series transformer network serves as the encoder in our training setup. The transformer, which is state-of-the-art and used in LLM (Large Language Model) and computer vision applications, is highly scalable, transferrable, and generalizable \cite{kenton2019bert}. Table \ref{tab:tfr_parameter} lists the proposed transformer encoder hyperparameter specifications in detail. The architecture is then supplemented with the projection MLP head. It features two fully connected layers and an input layer. Batch-normalization is applied to every fully connected layer, and the ReLU activation function is applied to every layer after batch-normalization, except the last layer. The next block in the architecture has prediction MLP. It has two fully connected layers. Batch-norm and ReLU are placed after the hidden layer of prediction MLP. The output layer, the final layer in the prediction MLP, has nodes equal to the number of output classes with Softmax activation. 

\subsubsection{Backbone training} \label{subsubsec:bbtrain}
The NNCLR framework-based transformer encoder network detailed in Section \ref{subsec:nnclrfw} is treated as a backbone network $\mathcal{B}$. It is trained with a dataset consisting of various mechanical faults and a healthy state of machines in both constant speed and varying speed conditions. A recently published experimental dataset \cite{jung2023vibration} is used for training the  $\mathcal{B}$. We have considered a healthy state, bearing fault (inner and outer race faults), and shaft misalignment faults from the dataset to train the backbone model $\mathcal{B}$. In this work, for the first time, we have introduced an active learning approach with contrastive learning to improve the model performance by enabling it to learn from the most informative samples and letting the model learn the precise decision boundaries. We have used entropy as an AL heuristic to determine informativeness within the unlabeled samples. Entropy is the most popular option for uncertainty-based querying approaches because it easily generalizes to probabilistic multi-class annotations and models more complex structural data.  Initially, $\mathcal{B}$ is trained with 15\% of the labeled samples. Now, to improve the performance of the model, we have actively sampled and labeled using the following method. 
We calculate the Shannon entropy of the predicted probability distribution over the classes for all the given samples. Based on this, we sample, say $2K$, samples that have the highest entropy. Now we label these $2K$ samples and calculate the KL-Divergence as follows: 
\begin{equation*}
    {D_{KL}}\left({q^{x}_{\mathcal{L}}} || {p^{x}_{\mathcal{M}}}\right) = \sum_{c} {q^{x}_{\mathcal{L}}(c)} \cdot \log{{q^{x}_{\mathcal{L}}(c)} \over {p^{x}_{\mathcal{M}}(c)}}
\end{equation*}
where $p^{x}_{\mathcal{M}}$ is the predicted probability distribution for sample $x$ by model $\mathcal{M}$ over the classes and $q^{x}_{\mathcal{L}}$ is the actual probability distribution to be given for the sample $x$ given its label class $\mathcal{L}$. It is defined as follows, 
\begin{equation*}
    q^{x}_{\mathcal{L}}(c) = 
    \begin{cases}
        1 & \text{if} \text{ c is } \mathcal{L}\\
        0 & \text{else}
    \end{cases}
\end{equation*}
Re-training the model with samples having highest Kl-Divergence shall help improve both accuracy as well as confidence. In this manner, we annotate an additional 5\% of the unlabeled data using AL and the model is subsequently re-trained. It is observed that the proposed training strategy enables us to achieve a classification accuracy of 94\%  with just 20\% labeled samples. On the other hand, to obtain that higher prediction accuracy with the random annotation method, it requires more than 60\% labeled samples, which clearly demonstrates the efficacy of the proposed approach.

\subsubsection{Target Tasks and Fine-tuning}

The target tasks considered in this work are divided into two groups to evaluate the performance of the foundational model $\mathcal{M}$ across fault cases and operating conditions (a) within the same machine, and (b) different machines. A hierarchical approach is followed to design the tasks in the first group. The initial few tasks aim to categorize the samples into healthy vs. faulty. Once the fault is detected, the subsequent target tasks are designed to diagnose the type, location, and severity of faults. Such a hierarchical approach enables a comprehensive fault diagnosis, offering detailed insights into the nature and location of potential faults. This is applied to both constant speed and varying speed conditions. The list of target tasks in the first group is tabulated in Table \ref{tab: target task}.  The second group of target tasks is designed to showcase the scalability and transferability of the proposed model $\mathcal{M}$. We have considered experimental vibration data from three different machines to evaluate our model performances across machines. We had frozen the transformer encoder from learning the new data representation. We made just the projection MLP head $\mathcal{N}$ trainable for performing downstream target tasks designed with the datsets for new machines. We have used only a small amount of labeled samples for fine-tuning the MLP head $\mathcal{N}$ for various downstream tasks, whereas the backbone network $\mathcal{B}$ was frozen with the weights that have been learned for the base machine dataset \cite{jung2023vibration}. The output layer of MLP head $\mathcal{N}$ was adjusted to the number of outputs based on the number of classes in the target tasks.

\begin{table}[]
		\centering 
		\caption{ Target Tasks performed with foundational model}
		\label{tab: target task}
\resizebox{\columnwidth}{!}{%
\begin{tabular}{cl}
\toprule
\multicolumn{1}{c}{S. No.} & \multicolumn{1}{c}{Target Tasks}
\\ 
\midrule
1                         & \begin{tabular}[c]{@{}l@{}}Fault Detection at constant speed: Classification - Healthy vs. Faulty\end{tabular}                                                                       \\
\midrule
2                         & \begin{tabular}[c]{@{}l@{}}Fault Diagnosis at constant speed: Classification - Bearing Fault vs \\Shaft Misalignment vs Rotor Imbalance\end{tabular}                 \\
\midrule
3                         & \begin{tabular}[c]{@{}l@{}}Bearing Fault Diagnosis at constant speed: Classification - Inner Race \\vs. Outer Race\end{tabular}                                         \\
\midrule
4                         & \begin{tabular}[c]{@{}l@{}}Fault Diagnosis at variable fault size: Classification - Healthy vs. Bearing \\Fault vs Shaft Misalignment vs Rotor Imbalance\end{tabular} \\
\midrule
5                         & \begin{tabular}[c]{@{}l@{}}Bearing Fault Diagnosis at variable fault size: Classification - Inner \\Race vs Outer Race\end{tabular}\\
\bottomrule
\end{tabular}
}
\end{table}

\begin{table}[]
		\centering 
		\caption{ The hyperparameters of the transformer encoder employed in the proposed model.}
		
\label{tab:tfr_parameter}
\begin{tabular}{lc}
\toprule 
Parameters                                                     & Value \\
\midrule
Input time series size                                       & 192 $\times$ 1 \\
Position encoding format                                       & 1D    \\
Number of stacked transformer blocks in the Transformer layer  & 4     \\
Number of heads in MSA                                          & 4     \\
Hidden layer size in a feed-forward network inside transformer   & 4     \\
Embedding dimension of the nonlinear transformation in MLP & 256   \\
Dropout probability                                            & 0     \\
Optimizer                                                      & Adam  \\
Initial Learning rate of the optimizer                     & 0.001\\
\bottomrule
\end{tabular}
\end{table}

\section{Results} \label{sec:results}

The backbone model is trained using the data collected from a bearing test rig at the Korea Advanced Institute of Science and Technology (KAIST). In order to verify and analyze the effectiveness of the proposed model in fault diagnosis across different machines, we have utilized datasets correspinding to three different machines: CWRU bearing dataset \cite{case2018case}, University of Ottawa dataset \cite{sehri2023university}, and HUST bearing dataset \cite{thuan2023hust}. In all the experiments, we shall consider results from our previous work in \cite{anbalagan2023foundational} as basline.

\subsection{Description of the datasets} \label{subsec:dataset}
The datasets used for designing the target tasks to evaluate the proposed framework on target tasks within the same machines as well as across different machines are described in this section.
\subsubsection{KAIST dataset}
 The data is collected from a bearing test rig at the Korea Advanced Institute of Science and Technology (KAIST). The dataset consists of two parts: the first part includes vibration, acoustic, temperature, and driving current data collected under varying load conditions. The data comprises measurements under three loading conditions: 0 Nm, 2 Nm and 4 Nm. The sampling frequency for vibration, temperature, and driving current data was 25.6 kHz. This dataset contains 120 seconds of vibration data in normal and 60 seconds in faulty states. The second part of the dataset focuses on vibration and current data acquired from the bearing faults at different locations such as inner-race, outer-race, and ball. These data were collected under continuously varying speed conditions by modifying the motor speed between 680 and 2460 RPM. For training the backbone in our work, we specifically consider the vibration data for bearing faults, shaft misalignment faults at a constant speed, and bearing faults at varying speeds.
We chose 60 seconds of data for each fault and normal condition at constant and variable speeds to maintain an equal amount of data in each class. More details about the dataset can refer to \cite{jung2023vibration}.
%The backbone model is trained by combining data from normal, inner race, outer race, shaft misalignment at constant speed, and normal, inner race, and outer race faults at variable speed conditions. 

\subsubsection{CWRU dataset}
The CWRU bearing dataset is the most popular in bearing fault diagnosis. The test bench comprises a 2-horsepower electric motor, a torque sensor, and a power dynamometer. To gather the vibration signals, accelerometers are mounted on the housings of the drive end and the fan end, respectively. The sampling frequency of drive-end fault data is 12 and 48 kHz, and the normal baseline and fan-end fault data are 12 kHz. The test bench uses deep groove ball bearings of the SKF-6205-2RS type to simulate pitting faults in bearings with diameters of 0.007 inches, 0.014 inches, 0.021 inches, and 0.028 inches that are machined using EDM (Electrical discharge machining) technology. The fault positions are located in the inner-race, outer-race, and ball locations. The bearing loads are 0, 1 HP, 2 HP, and 3 HP, respectively, corresponding to the different speeds. Outer race faults are divided into 6 o’clock, 3 o’clock, and 12 o’clock according to the location of the fault point. This study focused on the vibration data collected from the drive end with the sampling frequency at 48 kHz, and the fault diameter of 0.014 inches was considered. 

\subsubsection{HUST bearing dataset}
The HUST-bearing is a practical dataset for ball-bearing fault diagnosis released by the Hanoi University of Science and Technology. This dataset is especially advantageous because it contains fault signals from five bearings across different defects and working conditions. The data acquisition system includes a 1-HP induction motor, an accelerometer of PCB352C33, and a measurement module with torque and velocity sensors. This helps to evaluate the performance of the proposed method for various downstream tasks across different bearings or machines. We selected types 6205 and 6206 bearings for test analysis at 0W, 200W, and 400W loading conditions. Each type of bearing includes four health conditions: normal (N), inner race fault (I), outer race fault (O), and ball fault (B). The faults were generated using the wire-cutting method, which creates cracks of 0.2 mm. The accelerometer captured the vibration signals at a sampling rate of 51.2 kHz for 10 seconds. In this study, we considered the normal, inner, and outer race faults. The experimental evaluation of the proposed $\mathcal{M}$ to classify the selected two bearing type health state is demonstrated in Table \ref{tab: HUST}.

\subsubsection{University of Ottawa dataset}
The University of Ottawa Rolling-element Dataset - Vibration and Acoustic Faults under Constant Load and Speed circumstances (UORED-VAFCLS) test setup is used to acquire the data. The experimental setup consists of a single-phase motor mounted on a robust plate supported by anti-vibration mounts. A shaft adapter is used to connect the motor shaft, and an SKF E22206 spherical roller bearing is mounted on it to withstand the load applied by the cantilever beam. The motor operates at a constant speed of 1,750 RPM. The motor shaft is supported by two internal NSK 6203ZZ steel ball bearings. Notably, the first five tests employ 6203ZZ bearings, and the subsequent fifteen tests utilize FAFNIR 203KD bearings after a bearing model change. The data is collected with a sampling frequency of 42kHZ for 10 seconds, providing 420,000 data samples for each data sample. The dataset consists of healthy bearing, developing fault, and faulty bearing cases, which is useful for developing the fault diagnosis model sensitive to the incipient fault state and robust to developed fault conditions. In this experiment, we highlighted in Table\ref{tab: ottawa} that the proposed model is robust to classify both incipient and developed fault states with less labeled samples used for fine-tuning the $\mathcal{N}$.

\subsection{Efficiency}
Using the proposed approach, we have trained the $\mathcal{B}$ efficiently with less number of labeled samples. The contrastive learning method enables the transformer encoder to learn a better representation of the multiple fault cases from the KAIST dataset. We trained the backbone $\mathcal{B}$ with 15\% of labeled samples and attained 89\% classification accuracy. A better performance was observed after incorporating the AL query strategy to select the most informative samples and subsequently re-train the backbone using the proposed approach. When we implement the random sample selection method for training the $\mathcal{B}$, it takes 60\% labeled samples to reach 94\% classification accuracy. However, when the AL querying strategy is employed to select samples from an unlabeled pool of samples with just 20\% labels, we can reach 94\% accuracy. To evaluate the effectiveness of the trained backbone network $\mathcal{B}$, we have developed a list of hierarchical target tasks within the same machine data, and demonstrate that the proposed model $\mathcal{M}$ can perform better than baseline \cite{anbalagan2023foundational} in most of the target tasks presented in Table \ref{tab:tt compare}. The list of hierarchical target tasks framed from the KAIST dataset is shown in Table \ref{tab: target task}. The performance of the $\mathcal{B}$ in the hierarchical target tasks is shown in Table \ref{tab: TTASK1}. The results depict that the performance of the backbone $\mathcal{B}$ is higher due to the addition of the NNCLR framework with a transformer encoder network.

%%%%%%%%%%%%%%%%%%%%%%%%%%%%

\begin{table}[]
            \centering 
		\caption{ Comparison of backbone performance for target tasks within the same machine between the proposed model and baseline.}
		\label{tab:tt compare}

\begin{tabular}{ccc}
\toprule
\multicolumn{1}{l}{\multirow{2}{*}{Target Task}} & \multicolumn{2}{l}{Classification accuracy with 25\% labels}                \\
\cmidrule{2-3}
\multicolumn{1}{l}{}                             & \multicolumn{1}{l}{Proposed Model} & \multicolumn{1}{l}{Baseline Model} \\
\midrule
1                                                & 97.56                              & 95.33                              \\
2                                                & 99.62                              & 99.46                              \\
3                                                & 95.29                              & 99.33                              \\
4                                                & 95.07                              & 93.74                              \\
5                                                & 99.68                              & 99.39   \\
\bottomrule
\end{tabular}
\end{table}

%%%%%%%%%%%%%%%%%%%%%%%%%%%%
\begin{table}[]
		\centering 
		\caption{Fine-tuning the backbone for different target tasks within the same machine}
		\label{tab: TTASK1}
\begin{tabular}{crrrrr}
\toprule
\multirow{2}{*}{Target Tasks} & \multicolumn{5}{r}{Percentage of data used for fine-tuning} \\
\cmidrule{2-6}
                              & 5\%        & 10\%       & 15\%      & 20\%      & 25\%      \\
                              \midrule
1                             & 92.06      & 94.80      & 95.61     & 96.29     & 97.56     \\
2                             & 96.30      & 97.84     & 98.80     & 99.20     & 99.62       \\
3                             & 90.90      & 92.88      & 93.82     & 94.85     & 95.29       \\
4                             & 82.07      & 89.85      & 92.53     & 94.54     & 95.07     \\
5                             & 97.43     & 97.73     & 98.46     &  99.24      & 99.68  \\
\bottomrule
\end{tabular}
\end{table}

%\subsection{Robustness}

\subsection{Generalization}

Generalization, or the adaptability of a model, refers to its capacity to grasp and extrapolate from intricate combinations of more minor elements or attributes. This capacity is crucial for enabling the successful application of the model to novel contexts and environments, particularly in the context of fault diagnosis. The extent to which the model can grasp the interactions among diverse factors or components and discern their influence on the overall observed state or condition is the key factor that determines its level of generalizability. This, in turn, empowers the model to apply its knowledge effectively and make accurate predictions in unfamiliar and diverse settings. A generalized model can effectively handle variations in input and adapt to various environmental conditions. 
To express the generalization ability of the proposed model $\mathcal{M}$, we performed several downstream tasks in datsets corresponding to three different machines described in the section \ref{subsec:dataset}. For all downstream tasks, we fine-tune the backbone model by allowing only the MLP head to be trained while keeping the remainder of the encoder components frozen with the initial weights used during the training of the $\mathcal{B}$. The experimental results demonstrating the generalization ability of the proposed modeling approach across CWRU, Ottawa and HUST datasets are presented in Table \ref{tab: cwru}, Table \ref{tab: ottawa} and Table \ref{tab: HUST}, respectively. It can be observed that the predictive performance improves as more amount of data is used for fine-tuning the decision models. This is intuitive because a better representation is learned as the amount of input training data is increased.

% \usepackage{multirow}
% \usepackage{multirow}
% \usepackage{multirow}

% \usepackage{multirow}
\begin{table}[]
		\centering 
		\caption{Generalization: Fine-tuning the backbone for different target tasks in different machine (CWRU dataset)}
		\label{tab: cwru}
\begin{tabular}{cccc}
\toprule
\multirow{2}{*}{Loading condition} & \multicolumn{3}{r}{Data used for fine tuning} \\
\cmidrule{2-4}
                                   & 5\%                & 10\%               & 15\%              \\
                                   \midrule
0 HP                               & 94.70              & 97.62              & 98.76             \\
1 HP                               & 82.46              & 86.79              & 89.42             \\
2 HP                               & 82.14              & 87.11              & 90.89             \\
3 HP                               & 88.64              & 92.54              & 93.97          \\
\bottomrule
\end{tabular}
\end{table}

\begin{table}[]
		\centering 
		\caption{Generalization: Fine-tuning the backbone for different target tasks in different machines (Ottawa University dataset)}
		\label{tab: ottawa}
\begin{tabular}{cccc}
\toprule
\multirow{2}{*}{Fault type} & \multicolumn{3}{r}{Data used for fine tuning} \\
\cmidrule{2-4}
                                   & 5\%                & 10\%               & 15\%              \\
                                   \midrule
Developing fault                               & 93.95              & 96.22              & 97.27             \\
Bearing fault                               & 91.66              & 94.20              & 96.12             \\
Bearing fault                               & 94.96              & 96.55              & 96.81             \\
\bottomrule
\end{tabular}
\end{table}

% \usepackage{multirow}
\begin{table}[]
        \centering 
		\caption{Generalization: Fine-tuning the backbone for different target tasks in different machines (HUST dataset with two different types of bearings)}
		\label{tab: HUST}
\begin{tabular}{ccccccc}
\toprule
\multicolumn{1}{r}{\multirow{3}{*}{Loading condition}} & \multicolumn{6}{c}{Percentage of data used for fine-tuning}                                                                                                                                                              \\
\cmidrule{2-7}
\multicolumn{1}{r}{}                                   & \multicolumn{2}{c}{5\%}                                                & \multicolumn{2}{c}{10\%}                                               & \multicolumn{2}{c}{15\%}                                               \\
\cmidrule{2-7}
\multicolumn{1}{r}{}                                   & \multicolumn{1}{l}{Type 1} & \multicolumn{1}{l}{Type 2} & \multicolumn{1}{l}{Type 1} & \multicolumn{1}{l}{Type 2} & \multicolumn{1}{l}{Type 1} & \multicolumn{1}{l}{Type 2} \\
\midrule
0 W                                                    & 82.10                              & 89.20                               & 87.28                             & 92.15                              & 88.92                             & 93.17                              \\
200 W                                                  & 82.25                             & 89.06                              & 87.00                              & 91.95                              & 90.10                             & 93.79                              \\
400 W                                                  & 80.27                             & 85.77                              & 85.58                             & 89.48                              & 87.87                             & 92.56 \\    
\bottomrule
\end{tabular}
\end{table}

\section{Conclusion} \label{sec:conclusion}
 
This work proposes a novel active foundational model for fault diagnosis of the electrical machines by combining aspects of AL and contrastive SSL during backbone training. The proposed framework consists of two major componets: (i) constructing the backbone fault diagnosis model, and (ii) fine-tuning based on specific requirements of various target tasks. An inital model is trained using discrimination-based contrastive SSL technique, followed by re-training the same using informative samples selected via AL heuristics, namely, entropy and KL-Divergence. Once the backbone is trained, it is fine-tuned based on the requirements of different target tasks by updating the MLP projection head. The experimental evaluation has demonstrated the efficiency and generalizability of the proposed approach by achieving superior classification accuracy across several target tasks within the same machine as well as across different machines. The generalization capabilities of the proposed modeling approach makes it a valuable tool for real-world applications in fault diagnosis for electrical motors. The future extension of this work shall involve extensive testing across more target tasks and incorporating physics information within the foundational models.

% in order to overcome the limitations of obtaining meaningful labeled samples and harness the unlabeled samples in the learning process, and rely less on data augmentation technique, a new  was proposed based on the AL and nearest neighbor contrastive SSL. The training set samples are increased by labeling the most informative samples from the unlabeled pools. The informative samples from unlabeled samples are measured using entropy and KL divergence metric. Unlabeled samples with large uncertainties were selected for expert annotations to increase the size of the training set. Then, the transformer-based backbone network is constructed and trained using the proposed nearest neighbor contrastive self-supervised learning for 1D vibration signal. The developed backbone model can be fine-tuned for several target tasks within and across machines. The proposed approach was evaluated on three different induction motor bearing fault experimental datasets and performs better than existing fault diagnosis approaches with significantly less labeled samples. Future work will extend the proposed approach to diagnose faults across different types of machines.

\bibliography{afm_paper}

\end{document}